\documentclass[conference]{IEEEtran}
\ifCLASSINFOpdf
   \usepackage[pdftex]{graphicx}
\else
\fi
\ifCLASSOPTIONcompsoc
  \usepackage[caption=false,font=normalsize,labelfont=sf,textfont=sf]{subfig}
\else
  \usepackage[caption=false,font=footnotesize]{subfig}
\fi
\usepackage{booktabs}
\usepackage{amsmath}
\usepackage{amssymb}
\usepackage{paralist}
\usepackage{multirow}
\usepackage{xspace}
\usepackage{color}
\usepackage{xcolor}
\usepackage{ifthen}
\usepackage{url}
\usepackage{fancybox}
\usepackage{enumitem}
\usepackage{listings}
\usepackage{algpseudocode,algorithm,algorithmicx}
\usepackage{balance}
\usepackage{ifthen}
\usepackage{graphicx}
\usepackage{flushend}
\usepackage{tablefootnote}
\usepackage{todonotes}
\usepackage[utf8]{inputenc}

\definecolor{codegreen}{rgb}{0,0.6,0}
\definecolor{codegray}{rgb}{0.5,0.5,0.5}
\definecolor{codepurple}{rgb}{0.58,0,0.82}
\definecolor{backcolour}{rgb}{0.95,0.95,0.92}
 
\lstdefinestyle{mystyle}{
    backgroundcolor=\color{backcolour},   
    commentstyle=\color{codegreen},
    keywordstyle=\color{magenta},
    numberstyle=\tiny\color{codegray},
    stringstyle=\color{codepurple},
    basicstyle=\footnotesize,
    breakatwhitespace=false,         
    breaklines=true,                 
    captionpos=b,                    
    keepspaces=true,                 
    numbers=left,                    
    numbersep=5pt,                  
    showspaces=false,                
    showstringspaces=false,
    showtabs=false,                  
    tabsize=2
}
 
\usepackage{comment}

\algnewcommand\algorithmicforeach{\textbf{for each}}
\algdef{S}[FOR]{ForEach}[1]{\algorithmicforeach\ #1\ \algorithmicdo}

\begin{document}
\title{Using High-Rising Cities to Visualize \\ Performance in Real-Time}
%\title{Sky-Rise: A 3D Real-Time Visualization of \\ Software Performance}


\author{
\IEEEauthorblockN{Katsuya Ogami\IEEEauthorrefmark{1},
Raula Gaikovina Kula\IEEEauthorrefmark{1}, 
Hideaki Hata\IEEEauthorrefmark{1}, 
Takashi Ishio\IEEEauthorrefmark{1},
Kenichi Matsumoto\IEEEauthorrefmark{1}}
\IEEEauthorblockA{\IEEEauthorrefmark{1}Graduate School of Information Science\\
Nara Institute of Science and Technology,
Nara, Japan\\ Email: \{ogami.katsuya.ny7, raula-k, hata, ishio,  matumoto\}@is.naist.jp}
}

\maketitle

\begin{abstract}
For developers concerned with a performance drop or improvement in their software, a profiler allows a developer to quickly search and identify bottlenecks and leaks that consume much execution time. 
Non real-time profilers analyze the history of already executed stack traces, while a real-time profiler outputs the results concurrently with the execution of software, so users can know the results instantaneously. 
However, a real-time profiler risks providing overly large and complex outputs, which is difficult for developers to quickly analyze. 
In this paper, we visualize the performance data from a real-time profiler.
We visualize program execution as a three-dimensional (3D) city, representing the structure of the program as artifacts in a city (i.e., classes and packages expressed as buildings and districts) and their program executions expressed as the fluctuating height of artifacts. 
Through two case studies and using a prototype of our proposed visualization, we demonstrate how our visualization can easily identify performance issues such as a memory leak and compare performance changes between versions of a program.
A demonstration of the interactive features of our prototype is available at \url{https://youtu.be/eleVo19Hp4k}.
\end{abstract}

% \fi

%This is the proposed outline and presentation size of the paper. We will use this as a guide on how to space out the paper.
% INTRODUCTION -- Page 1 (add the introduce story, the case study and the final contributions of the paper)
% STATE OF THE ART -- Page 2 (half page) add the VisualVM and Jprofiler and the 
% VISUALIZATION DESIGN -- Page 2.5 
% CASE STUDY -- Page 4.5
% DISCUSSION -- Page 7
% RELATED WORK -- Page 8
% CONCLUSION and FINAL THOUGHTS -- Page 10
 
\IEEEpeerreviewmaketitle
\section{Introduction}
The first Lehman law of software evolution  \cite{Lehman:1996}  states that \textit{`a system must be continually adapted or it becomes progressively less satisfactory'}. 
Furthermore, Lehman's second law states that \textit{`an evolving system increases its complexity unless work is done to maintain or reduce it'}. 
As such, developers concerned with performance are required to constantly monitor the performance while evolving (i.e., making code changes) their applications. 
Developers utilize profilers as a means to quickly search and identify performance issues such as bottlenecks and memory leaks, which may consume the execution time of their applications.

State--of--the--art profilers profile application performance at a specific snapshot of an application.
According to an online report \cite{javaprofiler} in 2015, VirtualVM\footnote{\url{https://visualvm.github.io/}} and JProfiler\footnote{\url{https://www.ej-technologies.com/products/jprofiler/overview.html}} are comprehensive profilers for applications that run on the Java Virtual Machine (JVM).
%, as it brings numerous plugins and functionalities.
%VirtualVM has a feature of recording snapshots as seen in Figure \ref{fig:current} \footnote{Working with Snapshots \url{https://visualvm.java.net/snapshots.html}} and presents the information about your Java process performance. 
Both tools offer application developers the ability to properly profile their code execution, collect and browse thread dumps and heap dumps, while gathering various statistics about the internals of the JVM.

\begin{figure}[t]
\centering
\includegraphics[width=.8\linewidth]{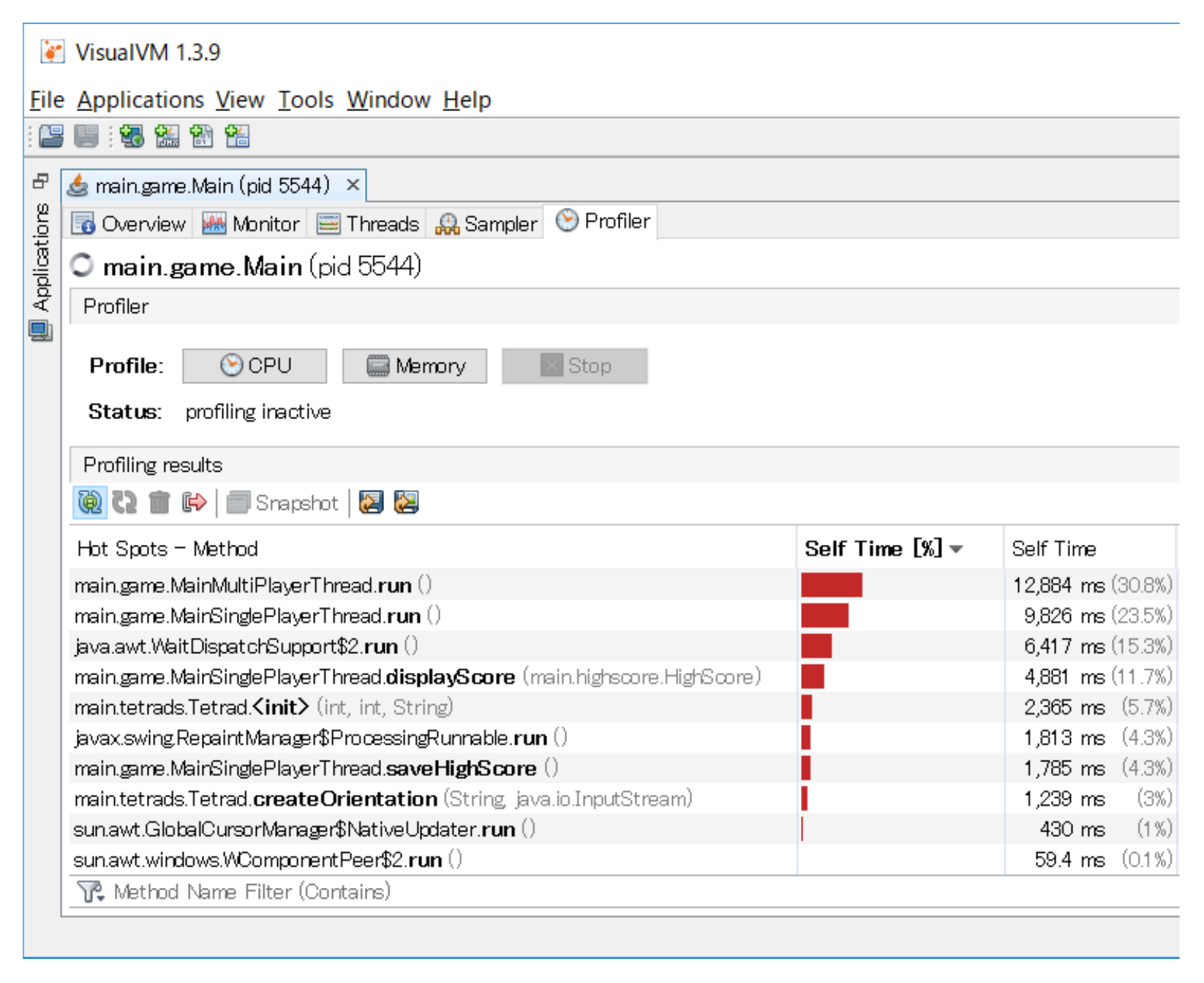}
\caption{A screenshot of VirtualVM showing program execution at a particular snapshot in time. Performance is visualized by a horizontal bar chart.}
    \label{fig:current}
\end{figure}

\begin{comment}
\begin{figure*}[t]
\centering
\subfloat[Snapshot of VirtualVM]{\includegraphics[width=.6\linewidth]{fig/current}%
\label{fig:current}}
\hfil
\subfloat[Real-time visualization while playing Tetris]{\includegraphics[width=.6\linewidth]{fig/proposal}%
\label{fig:proposal}}
\caption{Snapshot analysis to real-time profiling.}
\label{fig:motivation}
\end{figure*}
\end{comment}

Although comprehensive, we identify two drawbacks when using profilers such as VirtualVM and JProfiler. 
The first is the delay that exists between the execution of the program, logging of the execution stack traces and then profiling the results for a specific snapshot. 
Since profilers require time to process the information from an execution log, there is a delay between the actual running of the code and the analysis of its profile. 
The time taken to collect the data between executions can quickly become tedious and time-consuming for developers, especially if there are many different scenarios that the developer would like immediate feedback on performance while executing their application.
The second drawback is the design and presentation of user interfaces provided by these tools. 
As shown in Figure \ref{fig:current}, the typical screen presented by these tools are in the form of a horizontal bar chart. 
Although intuitive, in this example, other important information such as the relationship between the different components are not easily visible in a single interface.

\begin{figure*}[t]
\centering
\includegraphics[width=1\linewidth]{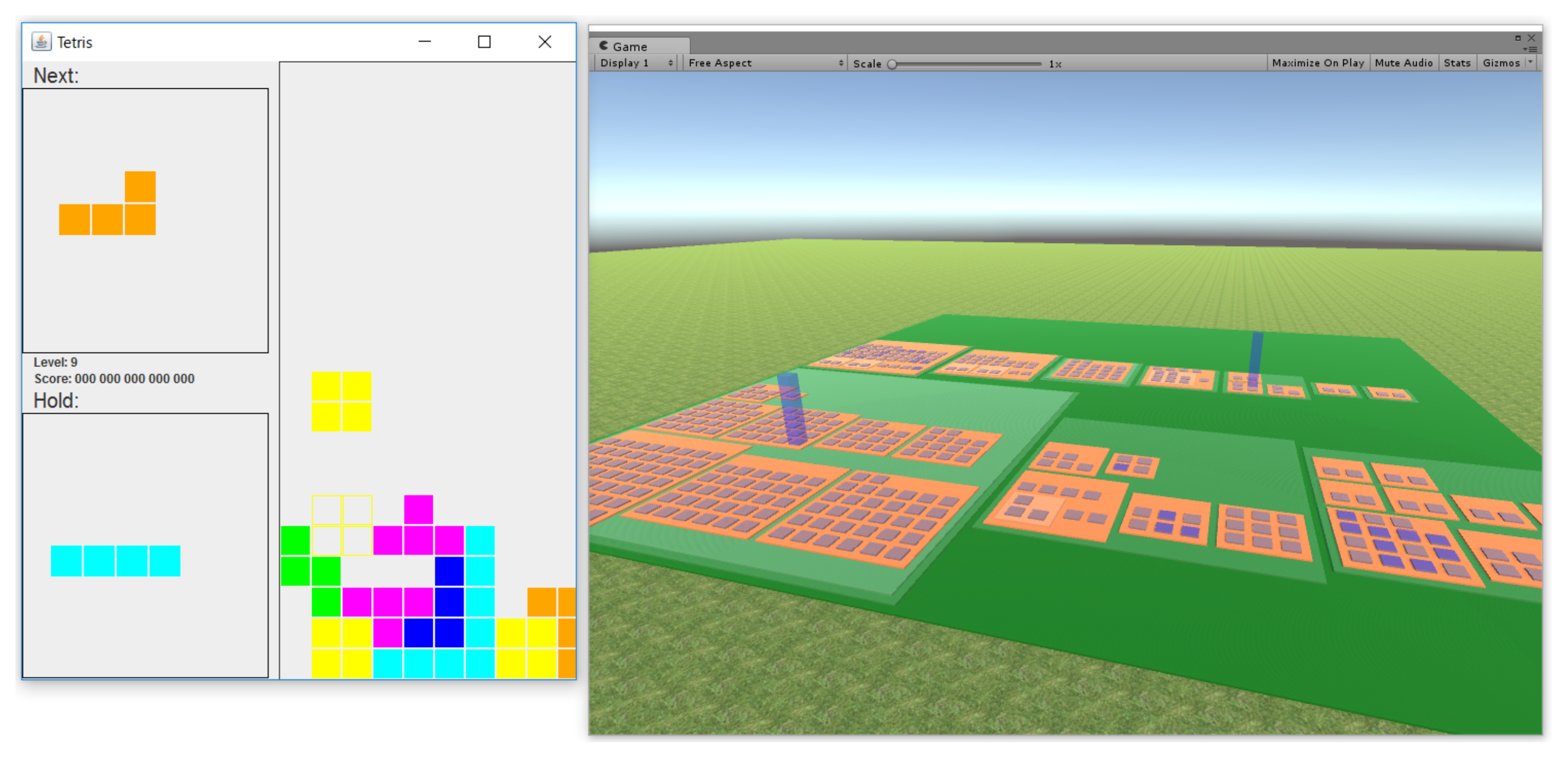}
\caption{Our prototype is a CodeCity visualization that shows real-time feedback on the performance of a\textit{`Tetris'} game.  
Fluctuations on the \textit{high-rise} buildings indicate performance executions based on the user interactions of a target program (i.e., user interaction of the  Graphical User Interface (GUI). )}
    \label{fig:proposal}
\end{figure*}

In this paper, we propose a three dimensional cities (commonly known as CodeCity) visualization to show performance in real-time to address these drawbacks.
As shown in Figure \ref{fig:proposal}, our proposed visualization (i) profiles in real-time, thus instantaneously updates the visualization while the application is running and (ii) depicts the underlying structure of the application with performance, thus providing a sense of the relationship between the different components in the application.
To evaluate the practicality and usefulness of our visualization, we present two case studies in which we (i) identify a potential performance issues (i.e., threads that are not properly killed) and (ii) depict performance during software  maintenance (i.e., performance and refactoring activities).

The paper is organized as follows.
Section \ref{sec:stateofart} presents the state--of--the--art and related work.
Section \ref{sec:visual} presents the visualization design and details while Sections \ref{sec:case} and \ref{sec:cases} presents the evaluation and case studies used to demonstrate the practicality and usefulness of our visualization.
We discuss aspects such as the generality, visual scalability, ease of use and practicallity in Section \ref{sec:discussion}.
We then finally conclude the paper in Section \ref{sec:conclusion}.

%\todo[inline, color=green]{Add the Introduction}
%\todo[inline, color=green!40]{Light green inline to-do}

\section{State of the Art}
\label{sec:stateofart}
In this section, we introduce the background and literature related to performance profilers.
We first introduce two performance profilers that are popular in industry. 
We then cover related literature in terms of the visualization of performance.

\subsection{Existing Tools in Practice}

VirtualVM is the state--of--the--art performance profiler that has a feature of recording snapshots, as seen in Figure \ref{fig:current}. 
According to the offical website, VisualVM is \textit{`an All-in-One Java Troubleshooting tool that takes advantage of several command line tools that are bundled with the Java Development Kit (JDK) distribution, and presents the information about the Java process performance'}.
Besides being a lightweight monitoring tool, VirtualVM offers application developers the ability to properly profile their code execution, collect and browse thread dumps and heap dumps, while gathering various statistics about the internals of the JVM.

%One of the most under-appreciated features is the ability to monitor MXBeans, management components that can show you thread pools usage, memory pools statistics, details about the frequency of garbage collection and so on. Typical memory problems like analyzing a heap dump to see what objects take the most of your JVM process memory. While other profilers might offer a more advanced approach to application profiling, VisualVM is the most appropriate first tier tool to analyze the performance of your code. Given that it’s bundled with your JDK distribution and is literally just one click away, it’s an invaluable tool! No wonder almost one in two developers is using it to either fully profile the application or in conjunction with other profilers to get a good initial impression using VisualVM and perhaps digging deeper with something heavier.

Another very popular performance profiling tool reported to be extensively used in industry is JProfiler. 
The official documentation state that \textit{`JProfiler is a comprehensive profiler for Java SE and Java EE applications with plugins for all major IDEs, which provides enhanced analysis of the collected profile data'}.
As with VirtualVM, JProfiler also provides CPU and memory profiling.
%Memory profiling with JProfiler can also be tuned to get more or less detail, for more data or to help to reduce the performance overhead.

As shown in Figure \ref{fig:proposal}, our visualization depicts the structure of software as different artifacts in CodeCity. 
We highlight two main differences between our proposed visualization against existing tools.
The first is the real-time feedback from the visualization. 
Both VirtualVM and JProfiler rely on logged data of executed program stack traces as either thread of heap dumps.
In our proposed method, we eliminate the process required to record and process the data collected in a stack trace. 
As shown in the figure, the visualization dynamically records as an application program is run.
This is especially advantageous for a user that requires instant feedback, possibly reducing the time needed to profile several snapshots of the application. 

The second difference between our proposed visualization and existing tools is inclusion of the program structure in the visualization, that provides insights into design aspects of the application. 
As shown in Figure \ref{fig:current}, the VirtualVM User Interface does not show the relationship between the called methods.
JProfiler does show a call graph view, where the methods are represented by colored rectangles that provide instant visual feedback about where the slow code resides in the method call chains, hence, making bottlenecks easier to find.
However, a large software system with hundreds of methods may generate call graph that is too complex to comprehend and aesthetically unpleasing for users.

%It can collect, analyze and render snapshots of the heap created with HPROF.

\subsection{Related Work}
%\section{Related Work}
\label{sec:related}

%We identified two types of related work: (i) work related to performance profiling and (ii) work related to visualization of program execution traces. 

% Performance analysis
The most related work is the `Performance Evolution Blueprint' \cite{6650523} and `Visualizing dynamic metrics with profiling blueprints' \cite{Bergel2010}. 
In these two works, Bergel and colleagues propose a profiling blueprint as a visualization that helps identify and remove performance bottlenecks.
Furthermore, the evolution blueprint evaluates performance during the evolution of the program.
Similarly, Bezemer et al, propose `Differential Frame Graphs' to understanding software performance regressions between two profiles \cite{7081872}.
Different to this work, our visualization considers the performance fluctuations that occur while running a single program version.

Many profiling tools and approaches utilize the call method relationship to describe the `design' of a program.
For example, a circular bundle \cite{Bundles2007} view visualizes the method call relationship.
Zinsight \cite{DePauw:2010:ZVA:1879211.1879233} visualizes event statistics and frequent method call patterns.
To reduce complexity, in this work, the circular bundle includes an interactive timeline, in which a user can customize and select a specific runtime snapshot size.

Much like VirtualVM and JProfiler, many profiling tools have been based on capture and profiling a snapshot of a running program \cite{CPE:CPE3156,Pauw:2001}.
However, there has been other work that have attempted close to real-time profiling.
In these works, the snapshots are referred to as phases, with phase detection approaches are proposed to split an execution trace into smaller phases corresponding to features in an execution.
For example, Watanabe et al. \cite{WatanabeWODA2008} proposed a technique to identify phases based on active objects.
Other work used key characteristics to highlight the phases.
Voigt et al. \cite{5306320} proposed a plot to visualize active objects in an execution trace.
Medini et al. \cite{Medini2012} proposed an effective algorithm that extract meaningful labels for detected phases.
For a real-time profiler, Reiss \cite{ReissWODA2005} proposed a technique to compare active methods between a pair of consecutive time slots.  
An important aspect of runtime monitoring and feedback is the overhead efficiency when profiling.
There exists work \cite{Ammons:1997} that exploits hardware performance counters to reduce runtime overhead for profiling.
Furthermore, to understand how time spent for an interaction of methods,
De Pauw et al. \cite{DePauw2006} proposed to use a customized UML sequence diagram whose time axis represents a real time.
Using detected phases, a developer can focus on a part of an execution trace. 
In our work, we create a fixed-time frame that provides a instantaneous feedback while a program in running.
Yamaguchi et al. \cite{1214461} proposed to visualize the behavior of distributed processes.  Their algorithm achieves an automatic, stable layout of rectangles representing hierarchical data elements.  Our method is dependent on the package structure of a program to achieve a stable layout. 

\begin{comment}
%It should be noted that visualization is not only the way to analyze a performance problem.
%Linhai Song, Shan Lu: Statistical debugging for real-world performance problems. OOPSLA 2014.
%A debugging approach (not a visualization) to analyze performance problems in an application.
%
%Memory Leak Analysis
%Guoqing Xu and Matthew Arnold and Nick Mitchell and Atanas Rountev and Gary Sevitsky: Go with the Flow: Profiling Copies To Find Runtime Bloat (PLDI 2009)

%\subsection{A 3D real-time analysis of Performance}
%\todo[inline, color=green]{Add the Limitations of existing tools and introduce our concept}
%One of the drawbacks of the existing tools, is the reliance on the execution stack traces. 
%This causes a delay between the execution and the generation of the profile. 

%As shown in Figure \ref{fig:motivation} most of the tools like Virtual VM show a simple barchart to depict the performance. 
%There is no information that pertains to the structure relation between the methods. 
\end{comment}

\section{Visualization Design}
\label{sec:visual}
\subsection{Cities Metaphor}
Wettel et. al \cite{Wettel:ICSE2011} introduce CodeCity as an integrated environment for software analysis, in which software systems are visualized as interactive, navigable 3D cities.
The classes are represented as buildings in the city, while their packages are depicted as the districts in which the buildings reside.
The visible properties of the city artifacts depict a set of chosen software metrics, as in the polymetric views of CodeCrawler.
The CodeCity metaphor has been the inspiration in numerous visualization scenarios of software. 
Early work include been Santos et al. \cite{EURECOM}, who proposed cities for visualizing information for network monitoring and later Panas et al. \cite{Parnas} proposed a similar idea for software production.  
Early researchers who actually implemented the city metaphor, \cite{Knight2000,Charters:2002} represented classes are districts
and the methods are buildings.
%More recently, similar work have proposed CodeCities to show....

We settled on a City metaphor, because is offers a clear notion of the locality and as explained by Wettel, \textit{`supports orientation and features a structural complexity that is not oversimplified'}. 
As shown by Wettel and Lanza \cite{Wettel:2008:softviz}, the CodeCity visualization can be leveraged to visually localize design problems.
We believe that this can be useful feature when profiling the performance of a program.
Moreover, we map the height of an artifact (i.e., such as a building) to provide a visual impression of the performance the artifact has in relation to the current executed functions of a program.
In our visualization, we conjecture that performance peaks and fluctuations are easily identified by the \textit{`high-risers'} in the city.

\subsection{Mapping Performance to the City}
The visualization is divided into two levels of analysis granularity.
Firstly, to intuitively express the structural properties of a software, we use the district localization.
The second level of mapping is the height of the buildings. 
The key point of the visualization is that the \textit{`high-rise'} buildings can be easily detected by the user.

\begin{figure}[t]
\centering
\includegraphics[width=.9\linewidth]{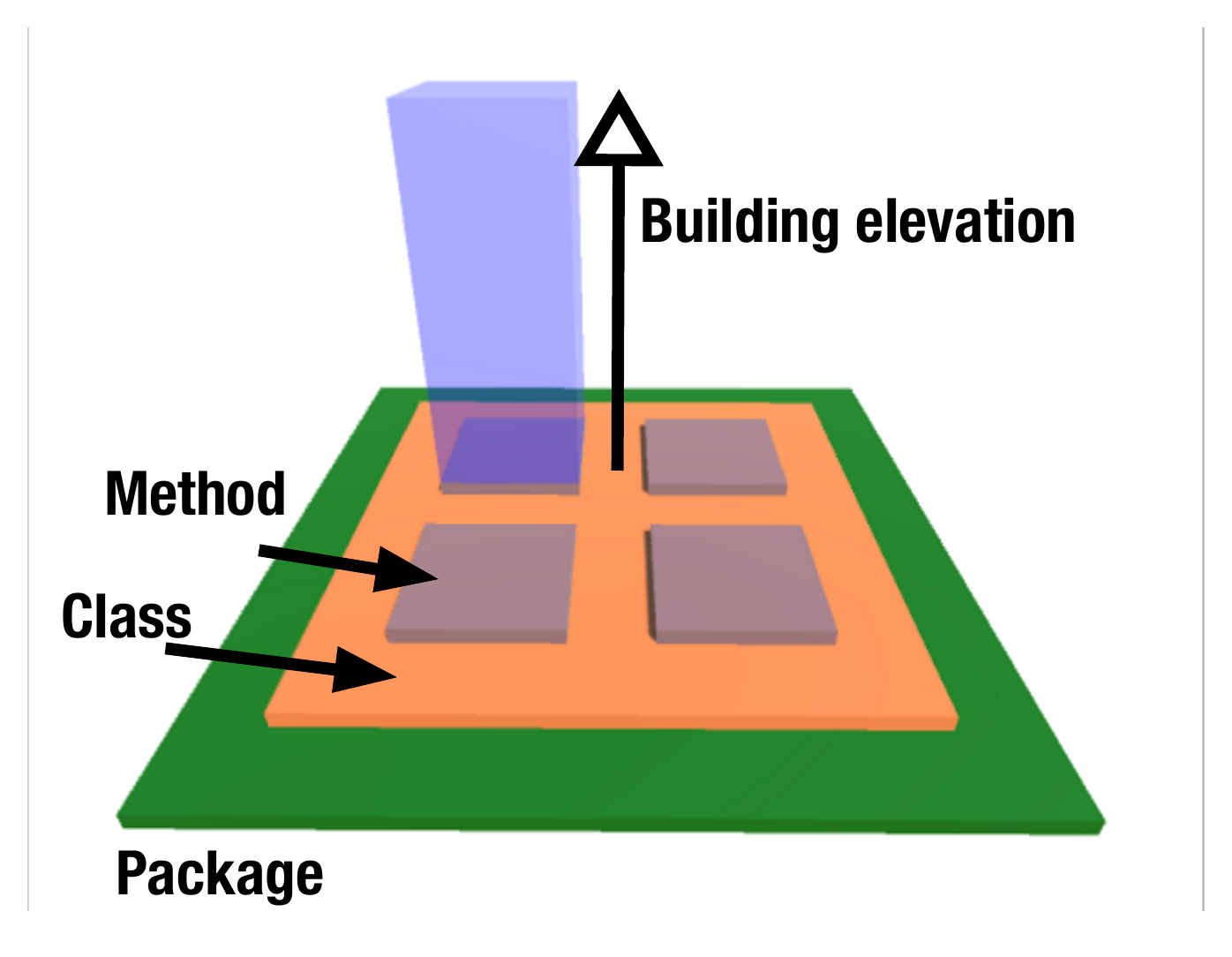}
\caption{Typical layout of a district in the city.}
    \label{fig:building}
\end{figure}

\begin{figure*}[t]
\centering
\includegraphics[width=.6\linewidth]{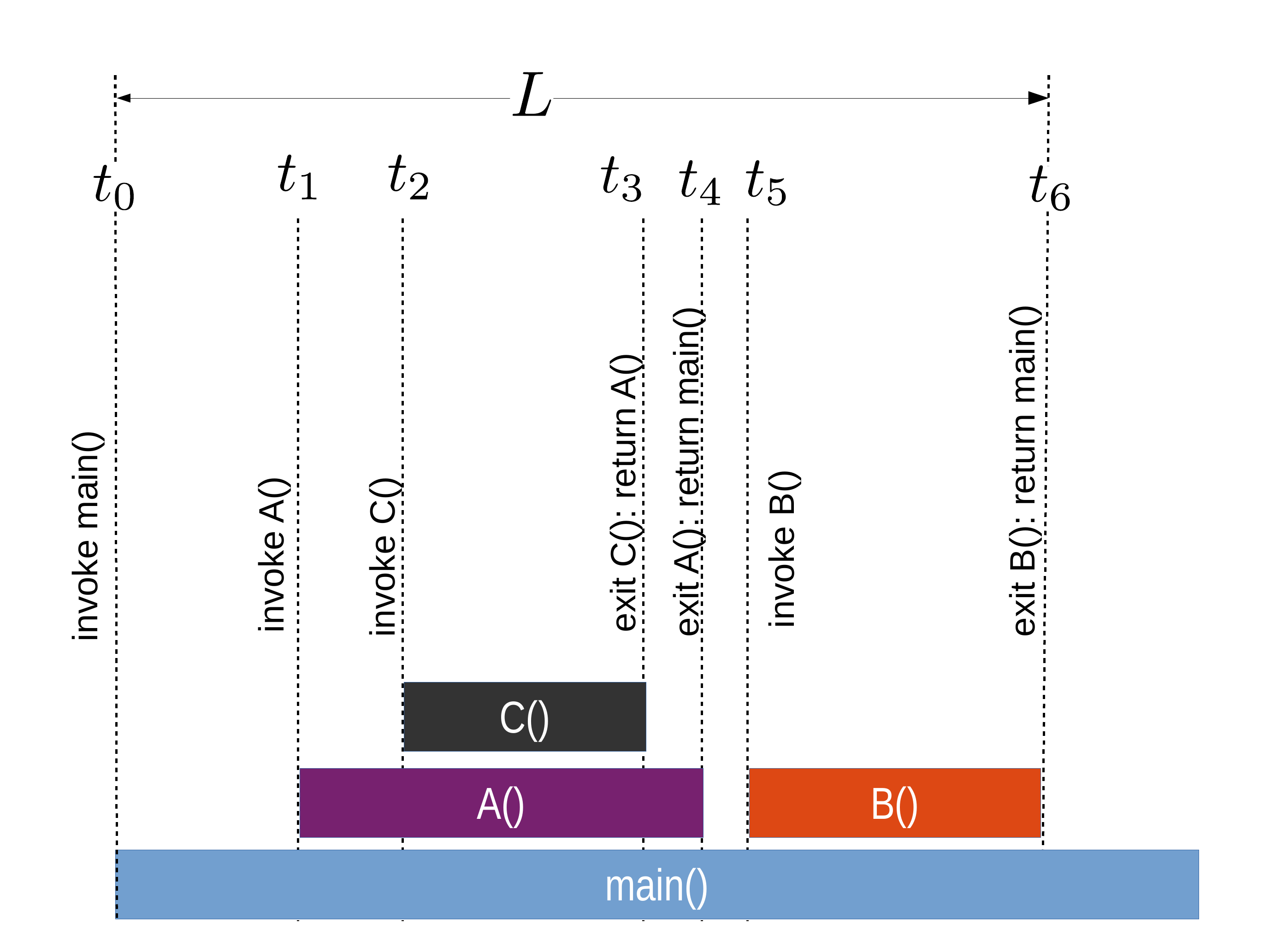}
\caption{Example of how we calculate the height of a building. In Figure \ref{fig:sequences}, we first show an example of an event $E_L$ where methods \texttt{main()}, \texttt{A()}, \texttt{C()} and \texttt{B()} are executed during time-frame $L$.}
    \label{fig:sequences}
\end{figure*}

\begin{comment}
\begin{figure*}[t]
\centering
\subfloat[Example showing the sequence of all events during (${E}_{L}$)]{\includegraphics[width=.5\linewidth]{fig/Seq}%
\label{fig:sequences}}
\hfil
\subfloat[Algorithm used to calculate building elevation ]{\includegraphics[width=.45\linewidth]{fig/elevation} 
\label{fig:elevate}}
\caption{Snapshot analysis to real-time profiling.}
\label{fig:formula}
\end{figure*}
\end{comment}

\subsubsection{District Design}
The district design has the main objective of depicting the structure of a software system.  
Figure \ref{fig:building} depicts the typical layout of the district.
As shown, every method is represented by standard uniform square-sized plot building.
Furthermore, methods are grouped according to the class (i.e,. classes are mapped to the same orange block).
Within in the same block, the grouped methods are ordered alphabetically by their names.  
Finally, classes are then grouped according to their packages. (i.e., common classes are grouped into the green district).
The visualization is also able to group according to sub-packages within a package. 
%This is mapped to within a district.

It is important to note that in the visualization that the packages and classes are ordered according to the number of methods contained. 
For example, the largest package with the most methods will be aligned at the bottom left of the visualization.
The rationale is that the users will be able to easily understand the performance in relation to structure of the program.

\begin{comment}
%\todo[inline, color=green]{For Fig. 2, Please change the execution time to elevation to be consistent with Fig 3b}
%\textbf{District Layout:}
%\textbf{Street Colors:}
\end{comment}

\subsubsection{Building Design}
%\textbf{Building Elevation:}

As shown in Figure \ref{fig:sequences}, our building elevation is based on a sequence of events within a fixed-time frame. 
Consider that $E$ represents all events during the execution of a program. 
Let $L$ be a fixed-time frame of interest.
Hence, we are interested in a subset of $E$, $E_{L}$ that contains all events that have been executed within the time-frame of $L$.
For example, in Figure \ref{fig:sequences}, methods \texttt{main()}, \texttt{A()}, \texttt{C()} and \texttt{B()} have been executed time-frame of $L$. 
Our algorithm uses sequences of executed time events (i.e., $t_{1}, t_{2} ... t_{i}$) where $i$ is the number of time events within $E_{L}$:

\begin{eqnarray*}
L = t_{i} - t_{0} \\
\end{eqnarray*}

As shown in Figure \ref{fig:sequences}, there are six events that occur during the $E_{L}$, hence, $L = (t_{6} - t_{0})$.
Each event consists of (i) timestamp, (ii) executed methods $M$ (iii) action.
The action is described as either a \texttt{invoke} or \texttt{exit and return} action.
These properties of an event are used to identify all events that occurred during the execution of method $m$.
For each method $m$, we now compute the accumulated execution time $Time(m)$ for the method as:

\begin{eqnarray*}
Time(m) = \sum_{\{i \in E_L| M(e_i) = m\}} t_{i+1} - t_{i} \\
\end{eqnarray*}

For example, for the method $A()$, the executed time is calculated as $Time(\texttt{A()}) = (t_2 - t_1) + (t_4-t_3)$.
%Note that $t_2, t_1, t_4~and~t_3$
Using $L$, we then normalize the executed time relative to the $E_{L}$.
Therefore, %as shown in Figure \ref{fig:elevate} 
we calculate for $m$:

\begin{eqnarray*}
Elevation(m) = \frac{Time(m)}{L} \\
\end{eqnarray*}

Using the example in Figure \ref{fig:sequences}, the height of the mapped building for method \texttt{A()} is calculated as follows:

\begin{eqnarray*}
Elevation(A()) = \frac{(t_2 - t_1) + (t_4-t_3)}{(t_{6} - t_{0})} \\
\end{eqnarray*}

where the elevation of method \texttt{A()} is calculated by using only the time when \texttt{A()} was invoked (i.e., we subtract the time when \texttt{C()} is invoked from when \texttt{A()} was invoked). We then normalize this over the total time-frame $L$.

\subsection{Technical Implementation}
A prototype of our visualization was developed using the game engine Unity\footnote{Unity at \url{https://unity3d.com/}} and implemented in the Java programming language. 
Unity is a 3D game engine that can render three-dimensional visualizations.  
For the prototype, our main consideration is how the executions was captured (through logging).
Specifically, we discuss some of the main technical considerations when implementing the logging of executions and the handling of Java threads.
%Related to the logging, we also discuss how we handle multiple thread executions.
%\textbf{Tools Used.} Our implementation used a Java program.
\begin{comment}
\todo[inline, color=green]{We should explain how logging is implemented. Does it store the recored time into a file, or directly transfer the data to the visualization tool (using a socket or something)?}
JVMがロードしたクラスのメソッドを走査して、メソッドの開始位置と、returnおよび例外throwのアセンブリコードの直前に、計測用メソッドを埋め込む。バイトコードの扱いにはASM(http://asm.ow2.org/)を利用した。得られた実行時間の情報は、一定時間ごとにまとめてJSON文字列に変換され、visualizerにTCPを用いて送信される（ファイルは生成しない）。JSONの生成にはjsonic(http://jsonic.osdn.jp/)を使用。すべてのメソッドにはIDが割り振られ、メソッド名などの情報は初回のみ送信される。
\end{comment}

\subsubsection{Logging of method executions}
To record execution time for each method, we use bytecode instrumentation.
We use the Java ASM\footnote{\url{http://asm.ow2.org/)}} library, which is a bytecode analyzer to inspect methods in any loaded classes of an executed Java program.
This is done by injecting logging code at the entry and the exit of the methods that invoke other methods (i.e., caller methods).
Specifically, the injection is performed just before the assembly code of a \texttt{return} or \texttt{exception throw}.
It is important to note that our approach of bytecode instrumentation is unable to inject logging code into methods protected by licenses or in cases of when the bytecode is unavailable (native methods).
Execution time of those methods are in the execution time of their callers.
Our prototype is able to exclude packages if required.
However, due to the design, the execution time of these excluded methods are still indirectly included in the caller method execution time.

The obtained sequences of events (i.e., $E_L$) and their execution times are then   transmitted in close to real-time batch via a Transmission Control Protocol (TCP) to our visualizer.
The visualizer is implemented with Unity (version 5.5.0f3)\footnote{\url{https://unity3d.com/}}. 
Each method is assigned a unique ID. 
All information such as the method names are sent in one transmission.
We set the fixed-time frame of $L$ at 3 seconds as it provides the most real-time experience, with trade-offs with the TCP and interface between the profiler and the user GUI of the executed program.

\begin{comment}
\todo[inline,color=green]{Does it inject logging code into libraries? (Is the time spent for a library method execution included in the caller's execution time?)}
 クラスパス以下にあるすべてのクラスが対象であるので、ライブラリのメソッドもプロファイリング対象に含まれる。また設定で、パッケージの名前を指定してプロファイリング対象から除外するように変更も可能である。除外されたパッケージのクラスのメソッドの実行時間は、callerメソッドの実行時間に含まれる。
 ツールは2つのモードを備えておりてのクラスをvisualizerで表示しておくモードと、JVMがロードしたクラスのみを表示してクラスロードのたびに配置を更新していくモードがある。
 java version "1.8.0_121", Unity version "5.5.0f3"
\end{comment}

\subsubsection{Handling of multi-threaded executions} 
Java uses threads to handle the concurrency that occurs within a program.
Our prototype is able to handle each thread independently as a sequence of events $E_L$.
In cases of multiple-threads accessing the same method, the longest executed time logged will be used as executed time of that method.
At this stage, it is possible for a lag in the logging process to delay the event sequence. 

%We use the longest time for each method among threads if they are executing the same method.
%The tool uses bytecode instrumentation to inject logging code.
%It records the entry and exit of methods.
%The injected logging code uses a network socket to the visualizer.
%Since a logging step may block the execution of the original process,
%the timer for logging is suspended during the execution of logging instructions.
%Here we should discuss the technical implementation of our tool, what it runs on.
%To record execution time for each method, we use bytecode instrumentation. (A pseudo code representing how logging code is embedded in a target program might be informative. )

\subsection{Real-time Interaction Properties}
A key highlight of our prototype is a dynamic analysis and feedback on performance in real-time (i.e., as a program is being executed). 
As shown in Figure \ref{fig:proposal}, we see that a target program (i.e., \textit{`tetris'} game application) is running simultaneously with our profiling visualization prototype.
Therefore, all game input such as Graphical User Interfaces (GUI) keystrokes and other user interactions are  visualized in real-time.

\begin{figure}[t]
\centering
\includegraphics[width=1\linewidth]{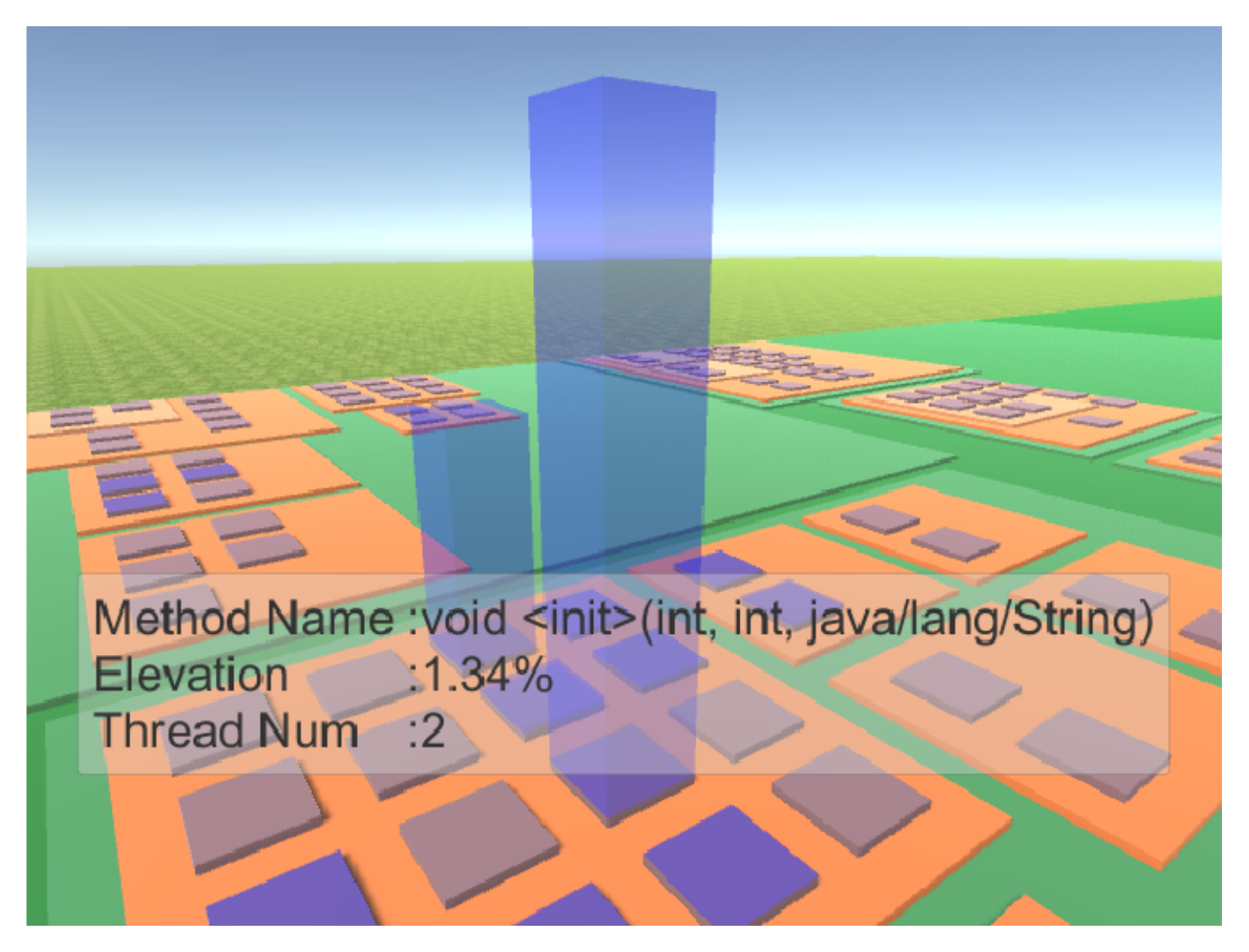}
\caption{Zoomed representation of our prototype. Upon the mouseover, we can see the (i) method name (ii) elevation and (iii) thread num of the method.}
\label{fig:zoom}
\end{figure}

Figure \ref{fig:zoom} presents a screenshot of a zoomed image of a method, in this case, a constructor of \texttt{void <init>(int, int, java,lang.String)}. 
As illustrated, the tooltip also shows the name of the method (i.e., Method Name), the elevation height (i.e., \%) and the number of threads (i.e., Thread Num) that have executed the method. 

To manage the layout of the city, we determined the sizes of the packages relative to the metric number of classes and methods.
Hence, the default aspect ratios for our package and class constructs.
Since the visualization is real-time, aesthetically, our intention is for simplicity so that users are quick to identify and focus on the changes in performance of the methods. 

An important aspect of our proposed visualization, which cannot be shown in the paper, is  the constant fluctuation of executed building heights during the real-time dynamic execution of the target program. 
The fluctuations are caused by both (i) the dynamic user interactions with the target program and (ii) the fixed time-frame $L$.
A video that demonstrates these dynamic fluctuating buildings is available at \url{https://youtu.be/eleVo19Hp4k}.

\section{Evaluation}
\label{sec:case}
%\subsection{Case Study Setup}
We study two real world cases to evaluate our prototype.
Our goal of the case studies is to demonstrate the practicality and usefulness of our proposed visualization.  
\begin{comment}
\begin{table}
\centering
\caption{Targeted Studied System Versions}
\label{tab:targetLib}
\begin{tabular}{@{}lrcccc@{}}
\toprule
\multicolumn{1}{l}{} & \multicolumn{1}{c}{Case Study One} & \multicolumn{1}{c}{Case Study Two} \\ 
\multicolumn{1}{l}{} & \multicolumn{1}{c}{Multi-Threading} & \multicolumn{1}{c}{Refactoring Activities} \\ 
Rationale & \\
\midrule \\
Commit Date & Oct 7, 2016 &　Oct 5, 2016\\
\midrule
Packages & 8 & 8\\
Classes & 24 & 24\\
Lines of Code & 1681 & 1682\\
\midrule
Changed files & & 22\\
Added Lines of Code & & 59\\
Deleted Lines of Code & & 58\\
\bottomrule
\end{tabular}
\end{table}
%\subsection{Case Setting: Visualization of Performance within a Game Application (Tetris)}
\end{comment}
The requirement criteria for our case study was as follows: (i) should be a Java  application that can be compiled and executed as a binary jar file, (ii) should have sufficient commit messages so that the history could be analyzed and (iii) 
have a sufficient code base that can be visualized as a city. 
Furthermore, it was preferable for the program to have a GUI to allow for more user interaction. 
In the end, we decided that a gaming application is suitable as most games have various GUI.
The classic game \texttt{`Tetris'} was selected due to its simple functionality and ease of keystrokes that would dynamically be displayed in our prototype.
After a systematic search (i.e., using the GitHub search function), the end result was a selection a Java version of Tetris\footnote{The website is available at \url{https://github.com/exal99/Tetris}} from five other tetris programs implemented in Java.
The main deciding factors included the code change history (i.e., 103 commits) and the program size (i.e., number of packages and lines of code).

\begin{figure*}[t]
\centering
\includegraphics[width=.7\linewidth]{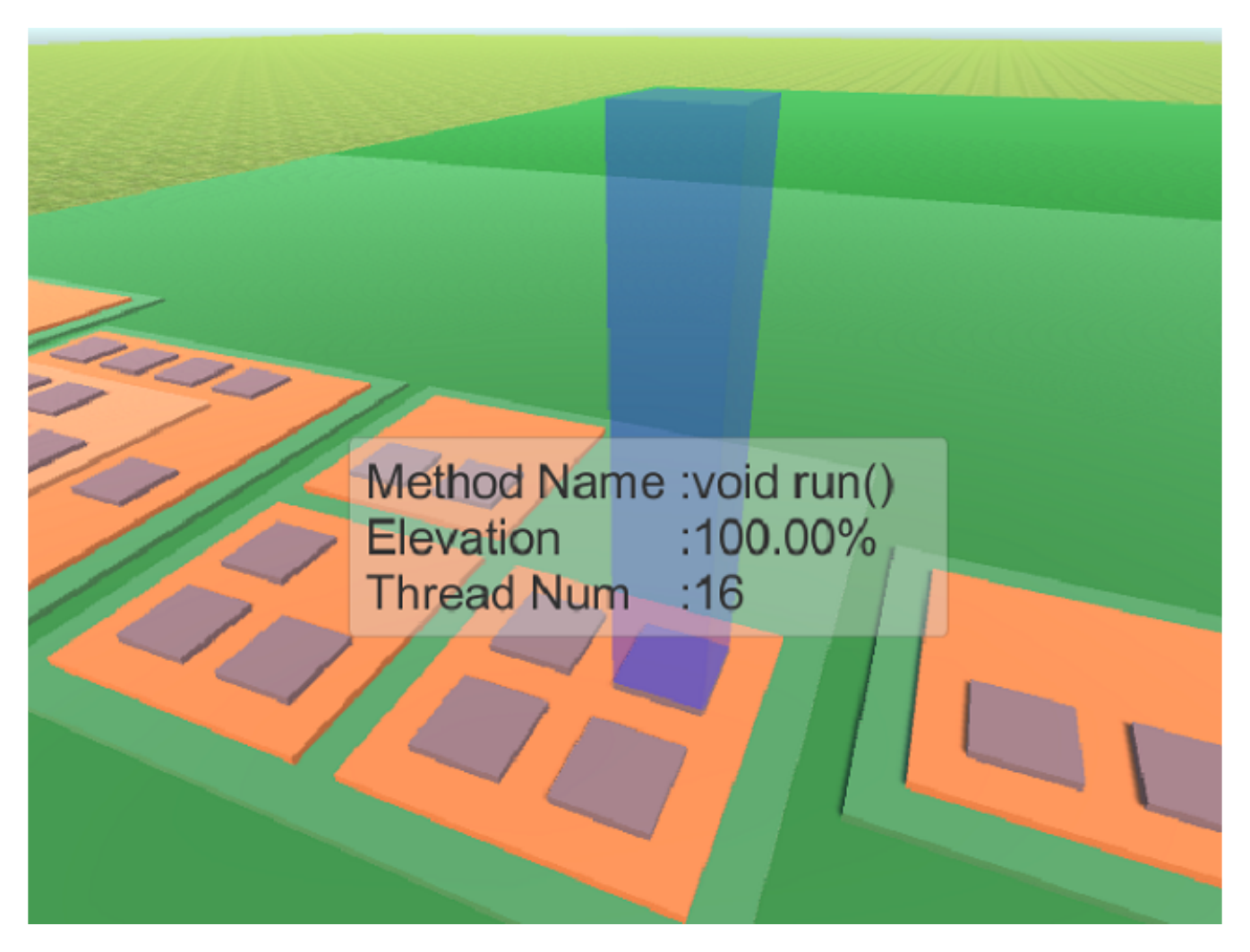}
\caption{A high-rise building indicates a performance issue. In this example, we find that the high riser is caused by a thread leak with multiple-threading. We find that the method invokes a new thread each time the user restarts for a new game.}
\label{fig:leak}
\end{figure*}

\begin{table}
\centering
\caption{Case One: Summary of program attributes during the identification of a Multi-Threading Leak}
\label{tab:case1}
\begin{tabular}{@{}lr@{}}
\toprule
Commit Date & Oct 7, 2016 \\
\midrule
\# of Packages & 8 \\
\# of Classes & 24 \\
\# of Lines of Code & 1681 \\
\bottomrule
\end{tabular}
\end{table}

\section{Case Studies}
\label{sec:cases}
We introduce two case studies that were used in our experiments.
The first case study is a demonstration of how the building height can be used to highlight sections in the code that are highly executed. 
We also would like to show how this simple but effective indicator of some potential performance issues in the source code.

The second case study is a demonstration of how we utilize the structure of the CodeCity metaphor to depict evolutionary changes in the target program. 
Specifically, we would like to show how the prototype is able identify structural changes that have over two versions of the target program. 
We also show and discuss how performance is affected by these code changes.

\subsection{Case One: Analysis of Performance with Multi-Threading}

The first case study demonstrates how our prototype is able to identify and monitor issues related to multi-threading and memory consumption in a software program.

Table \ref{tab:case1} describes the source code snapshot of the latest version of the target program (last modified Oct 2016).
The table shows that our target program consists of 8 packages, 24 classes and 1,681 lines of code. 
We find that the target program has a \texttt{run()} method is the interface that is intended to be executed by a thread.

Figure \ref{fig:leak} clearly shows a highly executed method (i.e., \texttt{run()}), with a high elevation (100\%). 
We notice that this building has no fluctuation in height and it constantly at 100\% height. 
Furthermore, the height of this building does not fluctuate, even when the user has ended a game and starts a new game. 

As shown in Figure \ref{fig:leak}, we can see that there is 16 threads created for the \texttt{run()} method.
This corresponds with the user interaction (i.e., user has restarted the game 16 times). 
Furthermore, the source code reveals that the \texttt{run()} method is comprised of a loop that creates a new thread each time a new game is created by a user. 
\pagebreak
\begin{lstlisting}[language=Java, caption=Code Snippet showing the \texttt{join()} method in the \texttt{MainSinglePlayerThread} class]
@Override
	public void run() {
		boolean running = true;
		while (isAlive() && !isInterrupted() && running) {
	 ...
		} else {
				saveHighScore();
				root.remove(graphics);
				root.add(new StartMenu(game, second, root, score, args));
				root.revalidate();
     ...
				try {
					join();
				} catch (InterruptedException e) {
					e.printStackTrace();
				}
     ...
		}
	}
 \end{lstlisting}
 
 \begin{figure*}[!t]
\centering
\subfloat[District design layout before refactoring operations were applied to some methods]{\includegraphics[width=.75\linewidth]{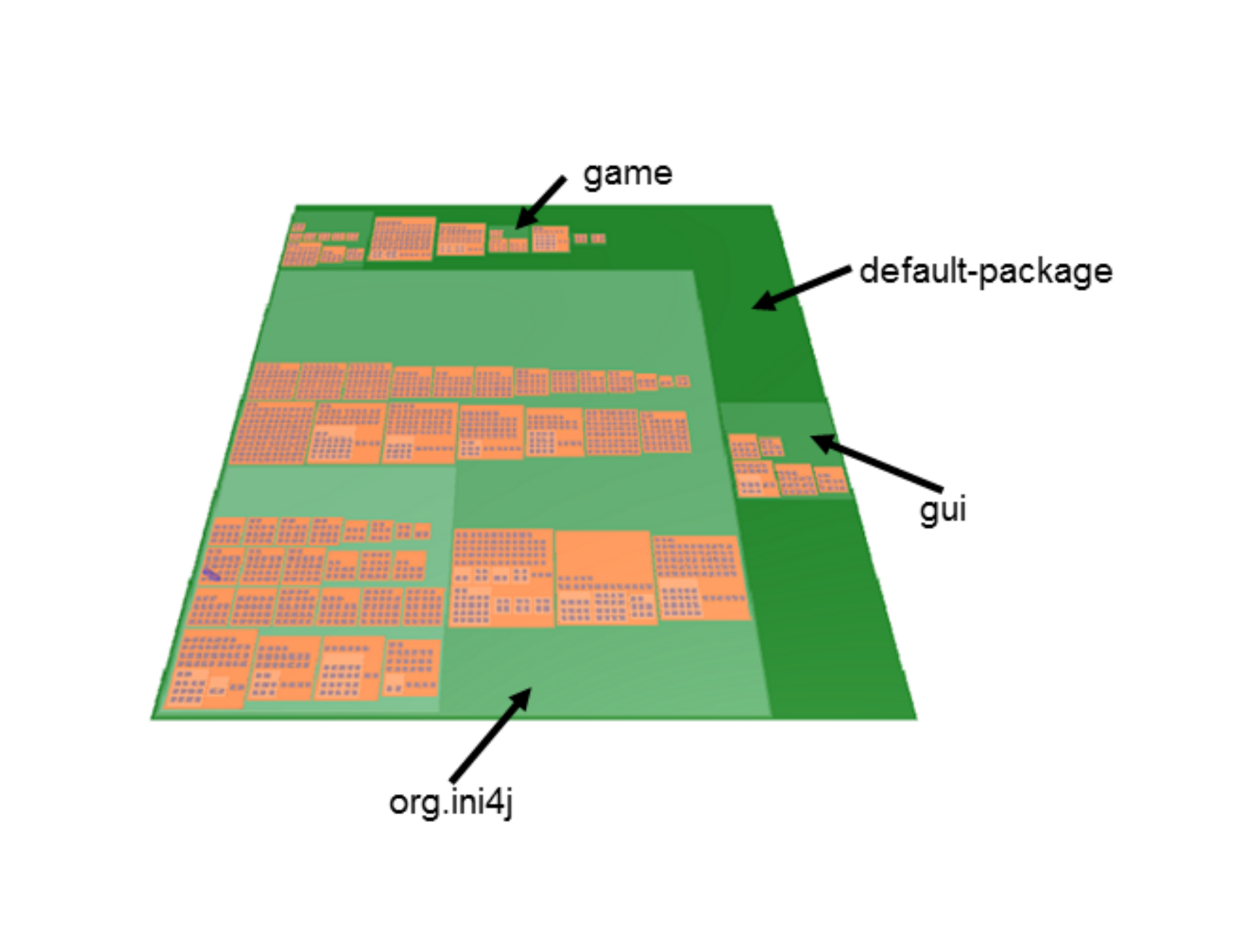}%
\label{fig:before}}
\hfil
\subfloat[District design layout after refactoring operations were applied to some methods]{\includegraphics[width=.75\linewidth]{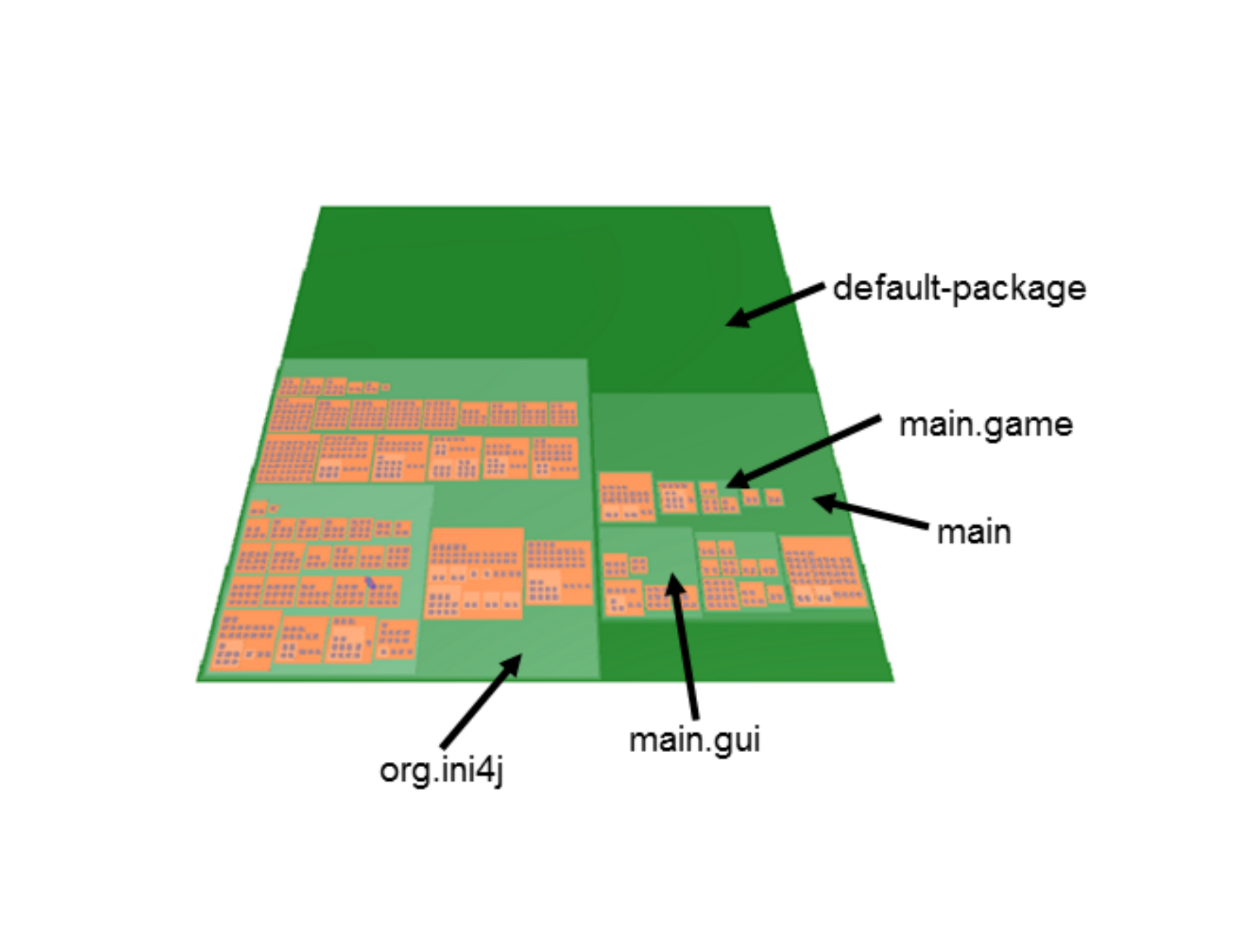}%
\label{fig:after}}
\caption{Evolution of the district structure of the city that reflect refactoring activities applied to the target program.}
\label{fig:refactoring}
\end{figure*}

 Listing 1 shows the snippet of the portion of the code that allows prior threads to remain alive (i.e., \texttt{MainSinglePlayerThread} class).
We are unsure of the developer's intention, however, we find that the \texttt{run()} method does not implement any management of the threading.
Potentially, this performance issue may result in a \texttt{OutOfMemoryExceptions} if these pool of threads are large enough, that will cause the application to crash. 
Due to the nature of the application (i.e., gaming application) and amount of memory consumed by each thread, we could assume that this issue may not have a significant impact on the usual performance of the application.
However, is a design concern as \textit{`programmers can typically get into trouble sizing data structures such as threads, understanding per-entry costs, and managing the lifetimes of these structures'} \cite{Mitchell:2007} . 

One possible fix to manage this concurrency issue, is to kill each thread every time a game has ended, especially for this single player mode. 
Subsequently, the program should start a new thread each time a new game is created.
To remedy this issue, we have contacted the developer with our recommended solution\footnote{We have raised a pull-request at \url{https://github.com/exal99/Tetris/pull/3}}.

\begin{table}[t]
\centering
\caption{Case Two: District Changes based on Refactoring Activities}
\label{tab:case2}
\begin{tabular}{@{}lcc@{}}
\toprule
& Before & After \\
Commit Date & Oct 5, 2016 & Oct 5, 2016 \\
\midrule
Classes & 24 & 24 \\
Lines of Code & 1681 & 1682 \\
\midrule
Changed Packages & \multicolumn{2}{c}{8} \\
Changed files & \multicolumn{2}{c}{22} \\
Added Lines of Code & \multicolumn{2}{c}{59} \\
Deleted Lines of Code & \multicolumn{2}{c}{58} \\
\bottomrule
\end{tabular}
\end{table}

\subsection{Case Two: Analysis of Performance in an Evolving City}
The second case study demonstrates how our prototype is able to model any structural changes as the program evolves.
Concretely, we show the performance when a set of refactoring operations have been applied to a newer version of the software. 
This example show be able to show that the refactorings have no effect on the performance of the application.

Table \ref{tab:case2} describes the source code snapshots between a refactoring commit on the source code (5th October 2016)\footnote{The commit is accessible at \url{https://github.com/exal99/Tetris/commit/925883f60f2dc2e57bd42ccf990637c89843ff8e}}.
We can see in the table that although the commit resulted in 1 additional line of code, there was code changes to 22 files we 59 added lines and 58 deletion. 
Upon the inspection of difference in the source code, we find that the commit included the integration of the \texttt{main} package, which involved importing (i.e., using the \textbf{import} keyword) new package structures. 
According to the commit message, the developers were \texttt{`Made all packegase subpackeges to prepare for the ai'}. 

Figure \ref{fig:refactoring} shows how our prototype able to detect and show these changes in the source code structure as landscapes changes of the district.
As shown in Figure \ref{fig:before}, before the refactoring, we find that there are nine districts in the city. 
The biggest district shown in the figure represents the third-party library, \texttt{Ini4j}\footnote{website at \url{http://ini4j.sourceforge.net/}}, which is a simple Java API for handling configuration files in Windows .ini format.
As highlighted in the figure, we observe that the classes related to the user interface (i.e., \texttt{gui} packages) are clearly in different locations compared to the game packages. 

Later in Figure \ref{fig:after}, after the refactoring, the structure of the city has changed.
Under closer inspection of the code, we observe that the developer had introduced a new main class (i.e., \texttt{main.java}) and \texttt{main} package into the program.
In the newer version, we find that all the \texttt{game} packages and classes have been merged into the the newer \texttt{main.game} packages. 
As shown in the Figures, these set of refactoring operations have no impact on performance.

Although this trivial example has no apparent impact on performance, it is a useful case to demonstrate how our prototype is able to depict changes in the program `design' structure between two versions of a program.

\section{Discussion}
\label{sec:discussion}
In this section, we discuss implications of our results. 
Specifically, we discussing the implications in terms of the generality, visual scalability, ease of use, practicality and potential future avenues such as additional scenarios for a user.

\subsection{Generality}
In our case study, we performed our experiments on a Java application. 
Currently our implementation is limited to executable Java bytecode binaries. 
However, we believe that this visualization can be extended to other programming languages. 

\subsection{Visual Scalability}
Indeed our visual design takes advantage of the CodeCity metaphor to depict program structure and identify performance issues.
However, the visualization as yet does not fully realize the other aesthetic features such as color and aspect ratio. 
In a real-time situation, however, we did consider whether or not other features would be more distracting and information overload. 
Therefore, our future plan would be to slowly add other useful elements to the visual design, while keeping the user engaged.

With the scalability, prior works show how the cities metaphor can be used to analyze massive software projects. 
In this example, we used a simple example with a single third-party library included.
However we believe that the visualization is scalable to larger code bases.
For future work, we would like to experiment with larger projects that have much more complex structures than our current case study.

\subsection{Ease of Use}
The CodeCity visualization has been popularly used in both the research community and the industry alike. 
With just elevation to represent performance, we conjecture that the visualization is both intuitive and very simple to use. 
To validate our claims, a user case study of our tool is needed and this is envisioned as a future extension of this work.

\subsection{Practicality}
Since current tools require a snapshot window, we conjecture that most profiling is specifically targets towards scripted sequences in a program.
We discovered that an advantages of the real-time profiling is the ability to run the program unscripted, thus potentially capturing these unexpected sequences while running a program.
This is especially beneficial for gaming applications that have complex GUI implementations.

Through case scenario executions of the prototype, we were able to demonstrate some of the useful applications of our visualization. 
However, in this study, we did not consider other considerations such as the overhead costs and delay with multi-threading.
This may become an issue with multiple threads with a large scale system.
As future work, we plan to measure the overheads using benchmarks such as the DaCapo Benchmark \cite{DaCapo:paper}. 
%How is the overhead of the tool?  We can measure overhead using a benchmark (e.g. DaCapo Benchmark), if we need to say the tool is low overhead.

\subsection{Additional Scenarios and Future Work}
In this paper, we present only two cases of how our visualization can be leveraged to show issues of (i) handling multi-threading and (ii) impact of refactoring activities. 
We believe that there is still many other scenarios that are needed to be tested. 
Below is are two scenarios that we would like to explore:
\begin{itemize}
%\item evaluating benefits of a real-time profiler.
\item evaluating scenarios that include large-scale target programs with complex GUI implementations.
%\item customizing visualization.
\item identification and differentiation of concurrent executions (i.e., multiple threads execution over the same method). 
A possibility would be tracking each individual thread and using visualization properties such as color.
\end{itemize}

Additionally, it would be beneficial to receive feedback from practitioners. 
Such feedback would allow us to make future improvements.
We should consider these future enhancements to our proposed solution as future work.
Compared to the related works that have used the CodeCity metaphor, our proposed visualization has not used many of the other aspects such as color, shape and area density properties. 
Therefore, as future work, we would like to investigate and utilize these properties to provide complementary information that a user may find useful.

\section{Conclusion}
\label{sec:conclusion}
In this paper, we present a visualization tool that is used to profile performance of a program in real-time that is based on the CodeCity visualization.
Our key is to demonstrate the usefulness and practicality of this visualization. 
Using a prototype, we illustrate through case studies of a real-world implementation of a gaming application to identify (i) analysis of a potential thread leakage and (ii) the performance in an evolving city.
For future work, we plan to further explore how the other city attributes could be utilized to enrich the profiling of performance for software applications.

\section*{Acknowledgment}
This work has been supported by JSPS KAKENHI Grant Number 16H05857 and JP26280021.

%\bibliographystyle{IEEEtran}
%\bibliography{IEEEabrv}

\begin{comment}
Notes
https://github.com/exal99/Tetris/commit/ddd57df59cedf726e5c94080f21b09856adb7c79
(1 changed files 33 additions) New Main file
https://github.com/exal99/Tetris/commit/47e2de115a54cdbc1795c15873149f1b9bb05c5d
(2 changed files 21 and 41) Moved Method
https://github.com/exal99/Tetris/commit/1d4437c93eb093f871654bc3daf068afc4820dfc
(1 file 21 add) Added Multiplayer
https://github.com/exal99/Tetris/commit/4974c88eae1dafb1b373c951ac5660260dfe8ae8
(7 files and big change: Working first prototype multiplayer)
that's all folks
\end{comment}

\end{document}